\documentstyle[prl,aps,preprint]{revtex}
\begin{document}
\draft
\title{Dual Description of Supergravity MacDowell-Mansouri Theory}
\author{H. Garc\'{\i}a-Compe\'an$^{a}$\thanks{%
Present Address: {\it School of Natural Sciences, Institute for Advanced
Study, Olden Lane, Princeton NJ 08540 USA}. E-mail: compean@sns.ias.edu}, J.
A. Nieto$^{b}$\thanks{%
E-mail: nieto@uas.uasnet.mx}, O. Obreg\'on$^{c}$\thanks{%
E-mail: octavio@ifug3.ugto.mx} and C. Ram\'{\i}rez$^d$\thanks{%
E-mail: cramirez@fcfm.buap.mx}}
\address{$^{a}$ {\it Departamento de F\'{\i}sica, Centro de Investigaci\'on y de
Estudios Avanzados del IPN}\\
P.O. Box 14-740, 07000, M\'exico D.F., M\'exico\\
$^b$ {\it Facultad de Ciencias F\'{\i}sico-Matem\'{a}ticas, Universidad
Aut\'{o}noma de Sinaloa}\\
C.P. 80000, Culiac\'{a}n Sinaloa, M\'{e}xico\\
$^c$ {\it Instituto de F\'{\i}sica de la Universidad de Guanajuato}\\
P.O. Box E-143, 37150, Le\'on Gto., M\'exico\\
$^d$ {\it Facultad de Ciencias F\'{\i}sico Matem\'aticas, Universidad
Aut\'onoma de Puebla}\\
P.O. Box 1364, 72000, Puebla, M\'exico}
\date{\today}
\maketitle

\begin{abstract}

In the context of field theory two elements seem to be necessary to search
for strong-weak coupling duality.  First, a gauge theory formulation and 
second, supersymmetry.  For gravitation these two elements are present in
 MacDowell-Mansouri supergravity.  The search for an ``effective
duality" in this theory  presents technical and conceptual problems
that we discuss.  Nevertheless, by means of a field theoretical approach,
which in the abelian case coincides with $S$-duality,  we exhibit a dual
theory, with inverted couplings. This results in a supersymmetric non-linear
sigma model of the Freedman-Townsend type.

\end{abstract}
\widetext


\newpage

\section{Introduction}

\setcounter{equation}{0}

Duality in supersymmetric gauge field theory and superstring theory is a
matter of great deal of recent investigation. It reveals
 new profound insights in the perturbative as well as in the
nonperturbative sectors of these theories (for recent reviews see
\cite{swreviews,sen}). In perturbative superstring theory, for instance,
$T$-duality and in general, mirror symmetry, gives a surprising new view
of space-time physics revealing the existence of a minimal length. It also
provides a powerful tool to study stringy phenomena of space-time physics,
such as the idea of worldsheet equivalence of non-smooth space-time
transitions whose prime example is, of course, the change in the
space-time topology. Furthermore, the strong-weak coupling duality or
$S$-duality has become a most important technique to study
non-perturbative aspects of field theory and string theory (see
\cite{swreviews,sen}). It turns out that $S$-duality is a fragile property
in the sense that the absence of supersymmetry can spoil it. In fact, there 
are known examples of non-supersymmetric field theories possesing the property 
of S-duality \cite{shamit}. However as far as duality is well understood in 
the presence of supersymmetry, we expect that in the case of MacDowell-Mansouri (MM) 
theory, supersymmetry will improve the S-dual description.
The absence of supersymmetry implies also the absence of the stability of
the spectra of masses under the renormalization group flow of BPS states
\cite{bps}.  In superstring theory, duality also shows the equivalence
among different types of perturbative string theories and indicates a
strong evidence of an emerging underlying theory known as $M$-theory
\cite{sen}.  On the other hand, for N=2 supersymmetric gauge theories in
four dimensions, Seiberg and Witten found that a strong-weak coupling
`effective duality' can be defined on its low energy effective theory for
the cases pure and with matter \cite{sw}.

Exploring analogies and generalizations of that genuine $S$-duality,
people obtained other kinds of ``dualities''. For instance, in
non-supersymmetric non-abelian gauge theory, it is possible to define a
``field theory duality"  by constructing explicitly the dual action to
Yang-Mills action with a theta term, following the usual
Ro$\check{c}$ek-Verlinde procedure \cite{lozano,ganor,mohammedi}.  The
resulting dual action is described by a non-linear ``massive"  sigma model
of the type worked out in \cite{west}. In particular, in \cite{ganor}, the
obtained (local gauge invariant) dual variables were used by the authors
to reproduce the large-$N$ limit of SU$(N)$ partition function gauge
theory in two-dimensional tori. In three dimensions, for SU(2), the dual
theory is shown to be very close to gravity theories and their moduli
space of solutions.

Following these ideas, in Refs. \cite{copr,cor} (and reviewed in
\cite{cupa})  it was shown that the MacDowell-Mansouri gauge theory
of gravity in four dimensions \cite{mm,pagels} does admit a dual
description in the sense of \cite{ganor}. This leads to a dual theory of
gravity with the structure of a generalized non-linear sigma model. 
It is important to mention that our approach does not concern the 
origin of the theory, it could be seen as an effective 
theory, for which renormalizability has not to be discussed. Nevertheless, the 
theory could be also considered as an elementary theory, and in this case its
quantization is an important issue.
Thus one would like to be able to find an ``effective gravitational
duality", in the sense that Seiberg and Witten have found for standard
gauge field theories.  It is not clear if such a program would work for
gravitation. However, it seems to us worthwhile to pursue this task.  In
the first place, we work with the MacDowell-Mansouri formulation, a gauge
theory of gravity.  The other element which seems to be necessary is
supersymmetry. So, a natural next step is to consider the MM-supergravity
version.  In this paper we work out its dual description in the same sense
of Refs.  \cite{copr,cor,cupa}.  We are aware that a more rigorous
definition of an ``effective gravitational duality" would need a way to
define a low energy effective theory \cite{sw}, among other requirements.
This task presents particular difficulties, by example, one needs to
preserve the gauge theory structure of the MM-theory when coupled to
matter \cite{wil}.  In order to find the dual description of the
MM-supergravity, in section II we present this theory and its (anti)
self-dual versions. In section III the dual action is obtained from a
parent action.  Section IV is devoted to discussion. An alternative
derivation of the dual action of MM-supergravity theory, showing
explicitly the formalism of calculus in superspace, is presented in the
appendix.

\section{MacDowell-Mansouri Supergravity}

The theory of MacDowell-Mansouri is a gauge theory in 3+1 dimensions, with
the anti-de Sitter SO(3,2) gauge group \cite{mm,pagels}. After breaking
the original gauge group to SO(3,1), the resulting gauge theory leads to
the Einstein-Hilbert action, cosmological constant term and Euler
topological invariant.

The supersymmetric version of MM theory can be constructed simply by
promoting the SO(3,2) gauge fields to those corresponding to the
supergroup Osp$(1|4)$. In particular the gauge potential ${\cal A}_\mu
^{~A}$ is a Osp$(1|4)$- Lie algebra valued potential. The corresponding
field strength is given by

\begin{equation}
{\cal F}_{\mu \nu }^{~A}=\partial _\mu {\cal A}_\nu ^{~A}-\partial
_\nu {\cal A}_\mu ^{~A}+\frac 12f_{~BC}^{A}{\cal A}_\mu ^{~B}{\cal
A}_\nu ^{~C},\label{efe}
\end{equation}
where $f_{~BC}^{A}$ are the structure constants of the super Lie algebra
Osp$(1|4)$. The field strength ${\cal F}_{\mu \nu }^{~A}$ can be decomposed
into three terms corresponding to the three generators $%
S_{A}=(S_{ab},P_a,Q_i) $ (with $P_a=S_{4a}$) of Osp$(1|4)$ as

\begin{equation}
{\cal F}_{\mu \nu }^{ab}=F_{\mu \nu }^{ab}+\Sigma _{\mu \nu }^{ab}+\Theta
_{\mu \nu }^{ab},\ \ \ {\cal F}_{\mu \nu }^i=F_{\mu \nu }^i+\Sigma _{\mu \nu
}^i,\ \ \ {\cal F}_{\mu \nu }^a=F_{\mu \nu }^a+\Sigma _{\mu \nu }^a,
\end{equation}
where

\begin{equation}
F_{\mu \nu }^{~ab}=\partial _\mu A_\nu ^{~ab}-\partial _\nu A_\mu ^{~ab}+%
\frac 12f_{cdef}^{ab}A_\mu ^{~cd}A_\nu ^{~ef},
\end{equation}

\begin{equation}
\Sigma _{\mu \nu }^{ab}=2f_{4c4d}^{ab}A_\mu ^{~4c}A_\nu ^{~4d},
\end{equation}
\begin{equation}
\Theta _{\mu \nu }^{ab}={\frac 12}f_{ij}^{ab}A_\mu ^{~i}A_\nu ^{~j},\
\end{equation}
\begin{equation}
\ \Sigma _{\mu \nu }^a={\frac 12}f_{ij}^{4a}A_\mu ^{~i}A_\nu ^{~j},
\end{equation}
\begin{equation}
\Sigma _{\mu \nu }^i=f_{4aj}^i(A_\mu ^{~4a}A_\nu ^{~j}-A_\nu ^{~4a}A_\mu
^{~j}),
\end{equation}
\begin{equation}
F_{\mu \nu }^{~i}=\partial _\mu A_\nu ^{~i}-\partial _\nu A_\mu ^i+\frac 12%
f_{cdj}^i(A_\mu ^{~cd}A_\nu ^{~j}- A_\nu^{~cd}A_\mu ^{~j}).
\end{equation}

The action proposed by MM to describe gravity and supergravity is
constructed only in terms of gauge fields, without the tetrad
or the spin 3/2 field,

\begin{equation}
S=\int d^4x\varepsilon ^{\mu \nu \rho \sigma }{\cal F}_{\mu \nu }^{~~A}{\cal %
F}_{\rho \sigma }^{~~B}M_{AB},
\end{equation}
where $M_{AB}$ is a generalization of the Levi-Civita symbol chosen by
MacDowell-Mansouri in the bosonic case  and which is defined by
\begin{equation}
M_{AB}=\pmatrix{ \varepsilon_{abcd}& 0 \cr 0&i(C\gamma_5)_{ij}},
\end{equation}
where $\gamma _5=i\gamma _0\gamma _1\gamma _2\gamma _3$, $\gamma _\mu$ are
the Dirac matrices and $C$ is the charge conjugation matrix which satisfies
$C \gamma_5 C^{-1} = \gamma^T_5$ and $C^T = -C$.  The above action can be
written as

\begin{equation}
S=\int d^4x\varepsilon ^{\mu \nu \rho \sigma }\bigg[{\cal F}_{\mu \nu }^{~ab}%
{\cal F}_{\rho \sigma }^{~cd}\varepsilon _{abcd}+i{\cal F}_{\mu \nu }^{~i}%
{\cal F}_{\rho \sigma }^{~j}(C\gamma _5)_{ij}\bigg].
\end{equation}
By identifying $A_\mu ^{~ab}\equiv \omega _\mu ^{~ab}$ with the spin
connection, $A_\mu ^{~4a}\equiv e_\mu ^{~a}$ with the tetrad and $A_\mu
^{~i}\equiv \psi _\mu ^{~i}$ with the gravitino, MM have shown that the action
(11) gives the gauge theory of N=1 supergravity, plus cosmological and
topological terms \cite{mm}.

We define the (anti) self-dual part of the field strength (1) as

\begin{equation}
{^{\pm }}{\cal F}_{\mu \nu }^{~~A}=\frac 12{^{\pm }}{\cal B}_B^A{\cal F}%
_{\mu \nu }^B,
\label{alfa}
\end{equation}
where ${^{\pm }}{\cal B}_B^A$ is given by

\begin{equation}
{^{\pm }}{\cal B}_B^A=\pmatrix{{^\pm}B^{ab}_{cd}&0\cr 0&{^\pm}B^i_j}=%
\pmatrix{\frac{1}{2}(\delta^{ab}_{~~cd} \mp i\varepsilon^{ab}_{~~cd}) &0\cr
0&  (1\pm \gamma_5)^i_j},
\label{beta}
\end{equation}
with $\delta _{~~cd}^{ab}=\delta _c^a\delta _d^b-\delta _d^a\delta _c^b$.
It is then not difficult to see that the self-dual action corresponding to (9)
 is given by

\begin{equation}
S^+=\int d^4x\varepsilon ^{\mu \nu \rho \sigma } {^+}{\cal F}_{\mu \nu }^{~~A}
 {^+}{\cal %
F}_{\rho \sigma }^{~~B}M_{AB},
\end{equation}
which in terms of the $M_{AB}$  components is

\begin{equation}
S^+=\int d^4x\varepsilon ^{\mu \nu \rho \sigma }\bigg[{^+}
{\cal F}_{\mu \nu }^{~ab}%
{^+}{\cal F}_{\rho \sigma }^{~cd}\varepsilon _{abcd}+i {^+}
{\cal F}_{\mu \nu }^{~i}%
{^+}{\cal F}_{\rho \sigma }^{~j}(C\gamma _5)_{ij}\bigg].
\end{equation}
This action is the one proposed in \cite{nieto2} as a self dual version of the
MM-supergravity gauge theory in terms only on the self-dual spin connection.
We will show that similar to Yang-Mills \cite{ganor,mohammedi} and
MacDowell-Mansouri theory \cite{cor}, the linear superposition of
(15) and the anti-self-dual corresponding  action, does admit
a dual description as well.

\section{Duality in  MacDowell-Mansouri Gauge Theory of Supergravity}

In this section we show that for a theory of gravity with the
structure of a gauge theory and which is also supersymmetric, it
is possible to find a dual description \cite{cor,cupa} given by a
supersymmetric non-linear sigma model.

We consider the following action consisting in the linear
superposition of the self-dual and anti-self-dual parts of (9)

\begin{equation}
I = \int d^4x\varepsilon^{\mu \nu \rho \sigma}\bigg( {^+}\tau {^+} {\cal F}%
^{A}_{\mu \nu} {^+}{\cal F}^{B}_{\rho \sigma}  - {^-}\tau {^-} {\cal F}%
^{A}_{\mu \nu} {^-}{\cal F}^{B}_{\rho \sigma} \bigg)
M_{AB}.\label{accion}
\end{equation}
This action can be rewritten as
\begin{equation}
\begin{array}{ll}
I = \int d^4x\varepsilon^{\mu \nu \rho \sigma}\bigg[ ({^+}\tau {^+} {\cal F}%
^{ab}_{\mu \nu} {^+}{\cal F}^{cd}_{\rho \sigma} - {^-}\tau {^-} {\cal F}%
^{ab}_{\mu \nu} {^-}{\cal F}^{cd}_{\rho \sigma} )
\varepsilon_{abcd} &  \\
+ i({^+}\tau {^+} {\cal F}^{i}_{\mu \nu} {^+}{\cal F}^{j}_{\rho \sigma} - {^-%
}\tau {^-} {\cal F}^{i}_{\mu \nu} {^-}{\cal F}^{j}_{\rho \sigma})
(C\gamma_5)_{ij}\bigg]. &
\end{array}
\end{equation}
Substituting (\ref{alfa}) and (\ref{beta}) into this action and using the
auxiliary formulas

\begin{equation}
\begin{array}{cc}
{^{\pm }}B_{\ \ ef}^{ab}{^{\pm }}B_{\ \
gh}^{cd}\varepsilon_{abcd}=\pm 2i B_{efgh} \\ {^{\pm }}B_r^i{^{\pm
}}B_s^j(C\gamma _5)_{ij}= \pm C_{rj} B_s^j, \label{formula}
\end{array}
\end{equation}
after some computations it is an easy matter to show that the
action (\ref{accion}) is equivalent to

\begin{equation}
I=\frac 12\int d^4x\varepsilon ^{\mu \nu \rho \sigma }\bigg[ ({^{+}}\tau -{%
^{-}}\tau ){\cal F}_{\mu \nu }^A{\cal F}_{\rho \sigma }^B+({^{+}}\tau +%
{^{-}}\tau )\tilde{{\cal F}}_{\rho \sigma }^B{\cal F}_{\mu \nu }^A%
\bigg] M_{AB}, \label{19}
\end{equation}
where $\tilde{{\cal F}}_{\mu \nu }^A$ consist of two parts
$(\tilde{{\cal F}} _{\mu \nu }^{ab},\tilde{{\cal F}}_{\mu \nu
}^i)$ defined by $\tilde{{\cal F}} _{\mu \nu }^{ab}={\frac
i2}\varepsilon _{~cd}^{ab}{\cal F}_{\mu \nu }^{cd}$ and
$\tilde{{\cal F}}_{\mu \nu }^i=\gamma _5{}_j^i{\cal F}_{\mu \nu
}^j.$ The first term in (\ref{19}) is the MM-supergravity (9).  In
\cite{cor} it was shown that the bosonic part of the second term
reduces to the Pontrjagin topological term.  The fermionic part of
this
 second term  is its corresponding supersymmetric
partner.

Now we will find a dual action to (\ref{accion}). In order to do
that, as usual, we propose a parent action. For the gauge theory
of supergravity of MM we propose this action as follows

\begin{equation}
L = \int d^4x\varepsilon^{\mu \nu \rho \sigma} \bigg(C_1 {^+}
{\cal G} ^{A}_{\mu \nu} {^+}{\cal G}^{B}_{\rho \sigma} + C_2 {^-}
{\cal G}^{A}_{\mu \nu} {^-}{\cal G}^{B}_{\rho \sigma} \\ + C_3
{^+} {\cal F}^{A}_{\mu \nu} {^+}{\cal G}^{B}_{\rho \sigma} + C_4
{^-} {\cal F}^{A}_{\mu \nu} {^-}{\cal G}^{B}_{\rho \sigma}\bigg)
M_{AB}
\label{preparent}
\end{equation}
where $C_i$ are constants and ${^\pm}{\cal G}$ are Osp$(1|4)$- Lie
algebra
 valued
Lagrange multiplier fields. Integrating out this action with
respect to ${^+}{\cal G},{^-}{\cal G}$ one
 easily recovers the action (\ref{accion}) with an appropriate choice of the coupling
constants $C_i$.

In order to get the dual theory we follow the reference \cite
{ganor,mohammedi}. Hence, one should start with the partition
function

\begin{equation}
Z= \int {\cal D} {^+} {\cal G} \, {\cal D} {^-} {\cal G} \, {\cal
D} {\cal A} \, exp( -L).
\label{feynman}
\end{equation}
The proposed parent action (\ref{preparent}) can be rewritten in terms of
the components of $M_{AB}$ in the form

\begin{equation}
\begin{array}{ll}
L = \int d^4x\varepsilon^{\mu \nu \rho \sigma}\bigg[ (
C_1{^+}{\cal G} ^{ab}_{\mu \nu} {^+}{\cal G}^{cd}_{\rho \sigma} +
C_2 {^-}{\cal G}^{ab}_{\mu \nu} {^-}{\cal G}^{cd}_{\rho \sigma})
\varepsilon_{abcd} &  \\ + i (C_1{^+}{\cal G}^{i}_{\mu \nu}
{^+}{\cal G}^{j}_{\rho \sigma} + C_2 {^-} {\cal G}^{i}_{\mu \nu}
{^-}{\cal G}^{j}_{\rho \sigma}) (C\gamma_5)_{ij} &
\\
+ (C_3 {^+}{\cal F}^{ab}_{\mu \nu} {^+}{\cal G}^{cd} _{\rho
\sigma} + C_4{^-} {\cal F}^{ab}_{\mu \nu} {^-}{\cal G}^{cd} _{\rho
\sigma}) \varepsilon_{abcd} &  \\ +i (C_3 {^+}{\cal F}^{i}_{\mu
\nu} {^+}{\cal G}^{j} _{\rho \sigma} + C_4 {^-} {\cal F}^{i}_{\mu
\nu} {^-}{\cal G}^{j} _{\rho \sigma}) (C\gamma_5)_{ij} \bigg]. &
\label{parent}
\end{array}
\end{equation}
In order to perform the Feynman integral we decompose this
functional integration measure $\int {\cal D}{\cal A}$ into its component
fields {\it i.e} $ \int {\cal D} {\cal A}^{A}_\mu = \int {\cal
D}A^{ab}_{\mu} {\cal D}A^i_{\mu} {\cal D}A^{4a}_{\mu}$. Thus we
define the dual action $L^*$ as follows

\begin{equation}
exp(-L^*) = \int {\cal D}A^{ab}_{\mu} {\cal D}A^i_{\mu}{\cal D}
A^{4a}_{\mu} exp\big( -L\big).\label{dual}
\label{componentes}
\end{equation}
The partition function written in terms of the dual action reads

\begin{equation}
Z= \int {\cal D} {^+} {\cal G}\, {\cal D} {^-} {\cal G} exp(-
L^*).
\end{equation}

Thus we first integrate out with respect to the field $A^{4a}_{\mu}$. Before
performing the functional integration it is convenient to take into account
that the
last two rows of action (\ref{parent}) can be rewritten as

\begin{equation}
\varepsilon^{\mu \nu \rho \sigma}(C_3 {^+}{\cal F}^{ab}_{\mu \nu} {^+}{\cal G%
}^{cd} _{\rho \sigma} + C_4{^-}{\cal F}^{ab}_{\mu \nu} {^-}{\cal G}^{cd}
_{\rho \sigma}) \varepsilon_{abcd} = -2i\varepsilon^{\mu \nu \rho \sigma}
{\cal F}^{ab}_{\mu \nu} (C_3 {^+}{\cal G}_{ab \rho \sigma} - C_4 {^-}{\cal G}%
_{ab \rho \sigma}),
\end{equation}
and
\begin{equation}
i \varepsilon^{\mu \nu \rho \sigma} (C_3 {^+}{\cal F}^{i}_{\mu \nu} {^+}%
{\cal G}^{j} _{\rho \sigma} + C_4 {^-}{\cal F}^{i}_{\mu \nu} {^-}{\cal G}%
^{j} _{\rho \sigma}) (C\gamma_5)_{ij} = i \varepsilon^{\mu \nu \rho \sigma}
{\cal F}^{i}_{\mu \nu} (C_3 {^+}{\cal G}^j_{~\rho \sigma} - C_4 {^-}{\cal G}%
_{~ \rho \sigma}^j) C_{ij}.
\end{equation}
Then the contribution to the field $A^{4a}_\mu$ comes from two pieces of the
parent action (\ref{parent})
\begin{equation}
\begin{array}{ll}
L = \int d^4x\varepsilon^{\mu \nu \rho \sigma} \bigg[\dots + 4i f^{ab}_{4c4d}
(C_3 {^+}{\cal G}_{\rho \sigma ab} - C_4 {^-}{\cal G}_{\rho \sigma ab})
A^{4c}_\mu A^{4d}_\nu & \\
+ 2i f^i_{4ak} A^k_{\nu} (C_3{^+}{\cal G}^j_{~\rho\sigma
} - C_4 {^-}{\cal G}^j_{~\rho \sigma}) A^{4a}_{\mu} C_{ij}+ \dots \bigg], &
\end{array}
\end{equation}
which can be written as an integral of the gaussian type. The result is

\begin{equation}
L^*_1 = \int d^4x \bigg[ \dots +  A^i_\mu [{\bf T}]^{\mu\nu}_{ij}
A^j_\nu + \dots\bigg] +  ln (\sqrt\pi det {\bf M}^{-\frac 12}),
\end{equation}
where
\begin{equation}
[{\bf T}]^{\mu\nu}_{ij} = - {\frac 14}[{\bf W}]^{\alpha \mu}_{ci} [{\bf M}%
^{-1}]^{cd}_{\alpha\beta} [{\bf W}]^{\beta \nu}_{dj},
\end{equation}
with
\begin{equation}
[{\bf W}]^{\mu\nu}_{aj} = 2i \varepsilon^{\mu \nu \rho \sigma} f^i_{4aj} (C_3%
{^+}{\cal G}_{\rho \sigma}^{~k} - C_4 {^-}{\cal G}_{\rho \sigma}^{~k}) C_{ik},
\end{equation}
and
\begin{equation}
[{\bf M}]^{\mu\nu}_{cd} = 4i \varepsilon^{\mu \nu \rho \sigma} f^{ab}_{4c4d}
(C_3 {^+}{\cal G}_{\rho \sigma ab} - C_4 {^-}{\cal G}_{\rho \sigma ab}).
\end{equation}

Now we perform the integral $\int {\cal D} A^i_\mu$ of fermionic variables
with $i$ being a Majorana index.
The terms in the Lagrangian $L^*_1$ that contribute to the fermionic path
integral are

\begin{equation}
L^*_1 = \int d^4x\varepsilon^{\mu \nu \rho \sigma} \bigg[ \dots + 2i
\Theta^{ab}_{\mu\nu} (C_3 {^+}{\cal G}_{\rho \sigma ab} - C_4 {^-}{\cal G}%
_{\rho \sigma ab}) + iF^{i}_{\mu\nu} (C_3 {^+}{\cal G}_{\rho \sigma}^{~j}
- C_4 {^-}{\cal G}_{\rho \sigma}^{~j}) C_{ij} + \dots \bigg].
\end{equation}
Contribution to the integral over $A^i_\mu$ from (29) and (33) can be
summarized as

\begin{equation}
L^*_1= \int d^4x \bigg[\dots + A^i_\mu[{\bf Z}]^{\mu\nu}_{ij} A^j_\nu + i
\varepsilon^{\mu \nu \rho \sigma}\bigg( f^i_{cdj}A^{cd}_\mu {\cal H}_{\rho
\sigma i} - 2  \partial_\mu {\cal H}_{\rho \sigma j}
\bigg) %
A^j_\nu +\dots \bigg],
\end{equation}
where
\begin{equation}
[{\bf Z}]^{\mu\nu}_{ij} = [{\bf T}]^{\mu\nu}_{ij} + [{\bf R}]^{\mu\nu}_{ij},
\end{equation}
${\cal H}_{\rho \sigma i}= (C_3 {^+}{\cal G}_{\rho \sigma}^j - C_4 {^-}{\cal %
G}_{\rho \sigma}^j) C_{ij}$ and

\begin{equation}
[{\bf R}]^{\mu\nu}_{ij} =i f^{ab}_{ij} \varepsilon^{\mu \nu \rho \sigma}
(C_3 {^+}{\cal G}_{\rho \sigma ab} - C_4 {^-}{\cal G}_{\rho \sigma ab}).
\end{equation}
The computation of the fermionic integral is given by

\begin{equation}
\begin{array}{ll}
L^*_2 = \int d^4x\bigg[\dots + {\frac 14} \varepsilon^{\mu \nu \rho \sigma} (
f^i_{cdj}A^{cd}_\mu {\cal H}_{\rho \sigma i} - 2  \partial_\mu
{\cal H}_{\rho \sigma j}) [{\bf Z}^{-1}]^{jk}_{\nu \beta}
\varepsilon^{\alpha \beta \gamma \delta} &  \\
( f^m_{cdk}A^{cd}_\alpha {\cal H}_{\gamma \delta m} + 2
\partial_\alpha {\cal H}_{\gamma \delta k}) + \dots \bigg] & \\
+ln (\sqrt{\pi} det {\bf Z}^{{\frac 12}}). &
\end{array}
\end{equation}

Finally we integrate out with respect $A^{ab}_\mu$. The terms of the action
that contribute to the integration are

\begin{equation}
\begin{array}{ll}
L^*_2=\int d^4x \bigg[\dots + {\frac 14} \varepsilon^{\mu\nu \rho\sigma}
f^i_{abj}
{\cal H}_{\rho \sigma i}[{\bf Z}^{-1}]^{jk} _{\nu \beta} \varepsilon^{\alpha
\beta \gamma \delta} f^m_{cdk}{\cal H}_{\gamma \delta m} A^{ab}_\mu
A^{cd}_\alpha &  \\
-  \varepsilon^{\mu\nu \rho\sigma} f^i_{cdj}{\cal H}_{\rho \sigma i}[{\bf Z}%
^{-1}]^{jk}_{\nu \beta} \varepsilon^{\alpha\beta \gamma \delta}
 \partial_\alpha {\cal H}_{\gamma \delta k} A^{cd}_\mu +
\dots \bigg]. &
\end{array}
\end{equation}
In addition we have the usual contribution coming form the interaction term

\begin{equation}
L^*_2 = \int d^4x\varepsilon^{\mu \nu \rho \sigma} \bigg[ \dots +
2iF^{ab}_{\mu\nu} (C_3 {^+}{\cal G}_{\rho \sigma ab} - C_4 {^-}{\cal G}%
_{\rho \sigma ab}) + \dots \bigg].
\end{equation}
Before performing the functional integration $\int {\cal D}A^{ab}_\mu$ with
respect to $A^{ab}_\mu$ it is convenient to make the following definitions

\begin{equation}
{\bf G}^{\mu \nu}_{cdef} = i \varepsilon^{\mu\nu \rho\sigma} f^{ab}_{cdef}
(C_3 {^+}{\cal G}_{\rho \sigma ab} - C_4 {^-}{\cal G}_{\rho \sigma ab}),
\end{equation}

\begin{equation}
{\bf F}^{\mu}_{ab} = 4i \varepsilon^{\mu\nu \rho\sigma} \partial_\nu (C_3 {^+%
}{\cal G}_{\rho \sigma ab} - C_4 {^-}{\cal G}_{\rho \sigma ab}),
\end{equation}

\begin{equation}
{\bf K}^{\mu \alpha}_{abcd} = {1\over 4}\varepsilon^{\mu\nu \rho\sigma} f^i_{abj}{\cal %
H}_{\rho \sigma i}[{\bf Z}^{-1}]^{jk}_{\nu \beta} \varepsilon^{\alpha\beta
\gamma \delta} f^m_{cdk} {\cal H}_{\gamma \delta m},
\end{equation}

\begin{equation}
{\bf V}^{\mu}_{cd} = - \varepsilon^{\mu\nu \rho\sigma} f^i_{cdj}{\cal H}%
_{\rho \sigma i} [{\bf Z}^{-1}]^{jk}_{\nu \beta}\varepsilon^{\alpha\beta
\gamma \delta}  \partial_\alpha {\cal H}_{\gamma \delta k}.
\end{equation}
The relevant Lagrangian is of the form

\begin{equation}
L^*_2 = \int d^4x \bigg[\dots + ({\bf K}^{\mu \nu}_{abcd} + {\bf G}^{\mu
\nu}_{abcd}) A^{ab}_\mu A^{cd}_\nu + ({\bf F}^{\mu}_{cd} + {\bf V}%
^{\mu}_{cd})A^{cd}_\mu + \dots\bigg].
\end{equation}
Finally we get the dual action in the sense of \cite{ganor,mohammedi}
 for
the MM gauge theory of supergravity, this is given by

\begin{equation}
\begin{array}{ll}
L^* = \int d^4x\bigg[ \varepsilon^{\mu \nu \rho \sigma} \bigg(( C_1{^+}{\cal %
G}^{ab}_{\mu \nu} {^+}{\cal G}^{cd}_{\rho \sigma} + C_2 {^-}{\cal G}%
^{ab}_{\mu \nu} {^-}{\cal G}^{cd}_{\rho \sigma}) \varepsilon_{abcd} &  \\
+ i (C_1{^+}{\cal G}^{i}_{\mu \nu} {^+}{\cal G}^{j}_{\rho \sigma} + C_2 {^-}%
{\cal G}^{i}_{\mu \nu} {^-}{\cal G}^{j}_{\rho \sigma}) (C \gamma_5)_{ij} &
\\
+4 \partial_\mu {\cal H}_{\rho \sigma j} [{\bf Z}%
^{-1}]^{jk}_{\nu \beta} \varepsilon^{\alpha \beta \gamma \delta}
 \partial_\alpha {\cal H}_{\gamma \delta k} \bigg) &  \\
-{\frac{1}{4}} ({\bf F}^{\mu}_{ab} + {\bf V}^{\mu}_{ab}) [({\bf K} + {\bf G}%
)^{-1}]_{\mu \nu}^{abcd} ({\bf F}^{\nu}_{cd} + {\bf V}^{\nu}_{cd})\bigg]  & \\
+ ln (\pi^{\frac 32} det {\bf M}^{-\frac 12} det {\bf Z}^{\frac 12} det
({\bf K}+ {\bf G})^{-\frac 12}). &
\end{array}
\label{tres}
\end{equation}
It should be remarked that this resulting supersymmetric non-linear
sigma model has the same structure of previous models considered in the
literature \cite{west}.

In the process to get the dual action (\ref{tres}) we have integrated out Feynman integral
(\ref{feynman}) through the explicit decomposition of the integration measure (\ref{componentes}).
This choice breaks explicitly its description in the superspace. It has not necessarily to be
so. It is interesting to see that the dual action (\ref{tres}) can
be also derived completely from the formalism of integration in supermanifolds
\cite{deWitt}. We leave the details of this derivation for the appendix A.

\section{Discussion}

In gauge field theory, $S$-duality arises naturally for abelian theories
\cite{witt}.  For non-abelian theories, Seiberg and Witten \cite{sw} have
shown that supersymmetry is very useful mainly due to the holomorphic
properties of the superpotential.  Thus, after breaking the gauge symmetry
an effective duality can be obtained.  For non-supersymmetric gauge
theories, a ``field theory duality" can be constructed
\cite{lozano,ganor,mohammedi}, which results in the usual S-duality for
the abelian case.  However, for non-abelian theories, this dual theory
turns out to be a kind of ``massive" non-linear sigma model \cite{west}.
Of course, $S$-duality is present in the superstring and M theory
approach, with deep non-perturbative consequences.  In \cite{hull},
``gravitational branes" which arise in type II superstrings and M theory
have been considered in the search for gravitational duality.

If one attempts to formulate a program to pursue $S$-duality for
gravitation, in the framework of field theories, it seems to us that one
should take into account the elements mentioned above for standard field
theories.

We need, first, a gauge theory of gravity.  This has been done already by
MM \cite{mm} and other authors \cite{pagels}.  Second, the theory should be 
supersymmetric.  Third, being the MM supergravity a gauge
theory, one would need to find matter couplings which should preserve this
supersymmetric gauge structure, in such a way that the tetrad and the
gravitino field do not appear in the action whole \cite{wil}.  Fourth, the
next step, is to break the symmetry in order to find an ``effective
gravitational duality".

The third and fourth steps present technical and conceptual challenges.
The coupling of matter in these kind of gauge theories of gravity is an
open problem \cite{wil}.  The symmetry breaking, consequently, has to be
understood, technically as well as conceptually.  Let us remember that we
are dealing with a gauge theory of (super)gravity.

In this work, following this program, we have been able to deal with a
gauge theory of gravity which is supersymmetric and we were able to find a
field theory duality in the sense of references
\cite{lozano,ganor,mohammedi}.  It is interesting to note that our
supergroup procedure could also be useful to extend the results for
standard Yang-Mills theories to their corresponding supersymmetric
versions.

The search to find an effective gravitational $S$-duality in the context
of field theory, requires further work and a deeper understanding of the
technical and conceptual issues.  Work in these aspects is in progress.

\centerline{\bf Acknowledgments}

This work was supported in part by CONACyT grants 3898P-E9608 and 28454-E.
One of us (H.G.-C.) would like to thank CONACyT for support
under the program {\it Programa de Posdoctorantes: Estancias Posdoctorales
en el Extranjero para Graduados en Instituciones Nacionales 1997-1998}
and the Institute for Advanced Study for its hospitality.

\newpage
\setcounter{section}{0}
\setcounter{subsection}{0}
\setcounter{equation}{0}
\renewcommand{\thesection}{Appendix \Alph{section}}
\renewcommand{\theequation}{\Alph{section}.\arabic{equation}}

\def\ce{{\cal E}}
\def\ota{\otimes_{\cal A}}
\def\bra#1{\left\langle #1\right|}
\def\ket#1{\left| #1\right\rangle}
\def\hs#1#2{\left\langle #1,#2\right\rangle}
\def\norm#1{{\Vert#1\Vert}}

\section{ Alternative Derivation of the Dual Action }

The purpose of this appendix is to provide an alternative derivation of MM-supergravity
dual action (45) from the formalism of integration in supermanifolds \cite{deWitt}.
First of all notice that the last two row of Eq. (23) can be written as

\begin{equation}
A_{\cal{F}}=-2i {\cal{F}}_{\mu\nu}^{ab} {\cal{H}}^{\mu\nu}_{ab}+
i{\cal{F}}_{\mu\nu}^{i} {\cal{H}}^{\mu\nu}_{i},
\end{equation}
where
${\cal H}^{\mu\nu}_{ab}= \varepsilon^{\mu \nu \rho \sigma}(C_3
{^+}{\cal G}_{\rho \sigma ab} - C_4 {^-}{\cal G}_{\rho\sigma
ab})$ and  ${\cal H}^{\mu\nu}_{i}= \varepsilon^{\mu \nu \rho
\sigma}(C_3 {^+}{\cal G}_{\rho \sigma i} - C_4 {^-}{\cal
G}_{\rho\sigma i}),$
with ${^\pm}{\cal G}_{\mu \nu i}=C_{ij} {^\pm}{\cal G}_{\mu \nu}^j.$

Taking into account the definition of the field strength ${\cal F}^A_{\mu \nu}$,
after partial integrations we get
\begin{equation}
\begin{array}{ll}
A_{\cal F}=&-K_{ab}^{\mu\nu} A_{\mu}^a A_{\nu}^b- 2
W_{ai}^{\mu\nu} A_{\mu}^a A_{\nu}^i-R_{ij}^{\mu\nu} A_{\mu}^i
A_{\nu}^j- G_{abcd}^{\mu\nu} A_{\mu}^{ab} A_{\nu}^{cd}\\
&+U_{abi}^{\mu\nu} A_{\mu}^{ab} A_{\nu}^i+4i \partial_\mu {\cal
H}^{\mu\nu}_{ab} A^{ab}_\nu- 2i \partial_\mu {\cal H}^{\mu\nu}_{i}
A^{i}_\nu,
\end{array}
\end{equation}
where $A^a_\mu= A^{4a}_\mu $ and $K_{ab}^{\mu\nu}=4 i
f^{cd}_{4a4b} {\cal H}^{\mu\nu}_{cd}$, $W_{ai}^{\mu\nu}=i
f^{j}_{4ai} {\cal H}^{\mu\nu}_j$, $R_{ij}^{\mu\nu}=i f^{ab}_{ij}
{\cal H}^{\mu\nu}_{ab}$, $G_{abcd}^{\mu\nu}=i f^{ef}_{abcd} {\cal
H}^{\mu\nu}_{ef}$ and $U_{abi}^{\mu\nu}=i f^{j}_{abi} {\cal
H}^{\mu\nu}_{j}$.

This quadratic form can be diagonalized by defining the $\tilde{A}$-variables

\begin{equation}
\begin{array}{ll}
A_{\cal F}=&-K_{ab}^{\mu\nu} \tilde{A}_\mu^a \tilde{A}_\nu^b+
{1\over 4} Z_{ij}^{\mu\nu} \tilde{A}_\mu^i \tilde{A}_\nu^j-
J_{abcd}^{\mu\nu} \tilde{A}_\mu^{ab} \tilde{A}_\nu^{cd}- 4 J^{-1\
abcd}_{~~\mu\nu} \partial_\rho {\cal H}^{\rho\mu}_{ab}
\partial_\sigma {\cal H}^{\sigma\nu}_{cd}\\
&+N^{ij}_{\mu\nu} \partial_\rho {\cal H}^{\rho\mu}_{i}
\partial_\sigma {\cal H}^{\sigma\nu}_{j}+
8 J^{-1\ abcd}_{~~\mu\nu} U^{\nu \theta}_{cdj} Z^{-1\
ji}_{~~\theta\tau} \partial_\rho {\cal H}^{\rho\mu}_{ab}
\partial_\sigma {\cal H}^{\sigma\tau}_{i},
\end{array}
\end{equation}
where

\begin{equation}
\begin{array}{ll}
&Z_{ij}^{\mu\nu}= 4(-R_{ij}^{\mu\nu}+ W_{ai}^{\rho\mu}
K_{~~\rho\sigma}^{-1\ ab} W_{bj}^{\sigma\nu}),\\
&J^{\mu\nu}_{abcd}=G_{abcd}^{\mu\nu}-U_{abi}^{\mu\rho} Z^{-1\
ij}_{~~\rho\sigma} U_{cdj}^{\nu\sigma},\\ &N^{ij}_{\mu\nu}=-4(
Z^{-1\ ij}_{\mu\nu}+ Z^{-1\ li}_{~~\sigma\mu}
U_{cdl}^{\theta\sigma} J^{-1\ abcd}_{~~\tau\theta}
U_{abk}^{\tau\rho} Z^{-1\ kj}_{~~\rho\nu} )
\end{array}
\end{equation}
and
\begin{equation}
\begin{array}{ll}
&\tilde{A}_\mu^a= A_\mu^a+K^{-1\ ab}_{~~\mu\rho}
W^{\rho\sigma}_{bi} A_\sigma^i,\\ &\tilde{A}_\mu^{i}=A_\mu^{i}+2
Z^{-1\ ij}_{~~\mu\nu} (2i \partial_\rho {\cal H}^{\rho\nu}_j-
U^{\rho\nu}_{abj} A_\rho^{ab}),\\
&\tilde{A}_\mu^{ab}=A_\mu^{ab}+2iJ^{-1\ abcd}_{~~\mu\nu}
(\partial_\rho {\cal H}^{\rho\nu}_{cd}- U^{\nu\rho}_{cdi} Z^{-1\
ij}_{\rho\sigma} \partial_\tau {\cal H}^{\tau\sigma}_{j}).
\label{variables}
\end{array}
\end{equation}
Now, if we define the matrix
\begin{equation}
{\cal M}=\left(
\begin{array}{cc}
G&U^T\\U&Z
\end{array}
\right) \label{matriz}
\end{equation}
whose inverse is given by (see for example \cite{deWitt})
\begin{equation}
\left(
\begin{array}{cc}
(1-G^{-1}U^TZ^{-1}U)^{-1}G^{-1}
&-(1-G^{-1}U^TZ^{-1}U)^{-1}G^{-1}U^{T}Z^{-1}\\ -(1- Z^{-1}U
G^{-1}U^T)^{-1}Z^{-1}U G^{-1}   &(1- Z^{-1}U G^{-1}U^T)^{-1}Z^{-1}
\end{array}
\right)
\end{equation}
and we take into account the identity
\begin{equation}
Z^{-1}U^TJ^{-1}U=1-(1- Z^{-1}U^TG^{-1}U)^{-1},
\end{equation}
we obtain, written in a matrix formulation

\begin{equation}
A_{\cal F}=-\tilde{A}^T
K\tilde{A}-\frac{1}{4}\tilde{A}^TZ\tilde{A}- \tilde{A}^T
J\tilde{A}+4{\cal M}^{-1\ AB}_{\mu\nu}
\partial_\rho {\cal H}_B^{\rho\nu} \partial_\sigma {\cal H}_A^{\sigma\mu} \label{28}
\end{equation}
In order to integrate the components of $A_\mu^A$, we must first
make the change of variables (\ref{variables}), it mixes the
variables among them and the corresponding jacobian should to be
taken into account. Due to the fact that bosonic and fermionic
variables come into play, the jacobian is given by a
superdeterminant, defined by \cite{deWitt}
\begin{equation}
Sdet\left(
\begin{array}{cc}
A&C\\D&B
\end{array}
\right)=\frac{ \det (A- C B^{-1} D)}{\det B},
\end{equation}
>From (\ref{variables}), it can be seen that the superdeterminant
of the matrix of the homogeneous part of the transformation is 1.
Thus, there is no contribution from the jacobian.

The integrations to be done are gaussian. Hence the result will be
the product of determinants $\det K^{-\frac{1}{2}} \det
J^{-\frac{1}{2}} \det Z^{\frac{1}{2}}$, where the last one has a
positive power because it comes from a fermionic integral. From
the definition of the matrix $J=G(1- G^{-1}UZ^{-1}U^T)$, we see
that the product of the last two determinants can be written as
the superdeterminant of the matrix (\ref{matriz}).

Thus we get that the dual action for MM-supergravity is given by

\begin{equation}
\begin{array}{ll}
L^* = \int d^4x\varepsilon^{\mu \nu \rho \sigma} \bigg(C_1 {^+}
{\cal G}^{A}_{\mu \nu} {^+}{\cal G}^{B}_{\rho \sigma} + C_2 {^-}
{\cal G}^{A}_{\mu\nu} {^-}{\cal G}^{B}_{\rho \sigma}\bigg) M_{AB}
+4{\cal M}^{-1\ AB}_{~~\mu\nu}
\partial_\rho {\cal H}_B^{\rho\nu} \partial_\sigma {\cal H}_A^{\sigma\mu}&\\
+\ln \left( \det K^{-\frac{1}{2}} \ \ {\rm Sdet}{\cal
M}^{-\frac{1}{2}}\right)&.
\end{array}
\end{equation}
Writing in components this action is, of course, equivalent to the action
(\ref{tres}).

\end{document}